\documentclass{ws-ijmpd}
\usepackage[super,compress]{cite}

\newcommand{\eqend}[1]{\,\mathrm{#1}}
\newcommand{\laplace}{\mathop{}\!\bigtriangleup}
\newcommand{\bx}{\textbf{x}}

\begin{document}

\markboth{N. Rendell}
{Large-distance behaviour of the massless vector propagator}

%
\catchline{}{}{}{}{}
%

\title{Large-distance behaviour of the massless vector two-point function in de Sitter spacetime}

\author{Nicola Rendell}

\address{Department of Mathematics, University of York, \\ 
Heslington, York, YO10 5DD, United Kingdom \\
nlr512@york.ac.uk}

\maketitle

\begin{history}
\received{29 January 2018}
\revised{2 March 2018}
\accepted{9 March 2018}
\end{history}

\begin{abstract}
We study the long-distance behaviour of the massless vector propagator in \(n\)-dimensional de Sitter spacetime, where \(n \geq 4\). Specifically, we consider the massless limit of the vector propagator in the Stueckelberg theory, which is an extension of Proca theory, with an additional gauge-fixing term. We work to leading order in the de Sitter-invariant distance \(Z\) to show that, in the large \(Z\) limit, this propagator tends to a gauge dependent constant, where the gauge worked in is described by the Stueckelberg parameter \(\xi\). In the Landau gauge, where \(\xi=0\), this constant is found to be 0. This result is in agreement with the 4 dimensional case discussed by Youssef. 
\end{abstract}


\ccode{PACS numbers: 04.62.+v, 11.15.-q, 14.70.-e}


\section{Introduction}

In its early stages, our universe is believed to have undergone an inflationary epoch \cite{InflationG} \cite{InflationS} \cite{InflationL} \cite{InflationAS} . As de Sitter spacetime gives an approximation of this inflationary epoch, it is of interest to study quantum field theories in this spacetime.

We study the large-distance behaviour of the massless vector propagator. To this end, we will consider the massless limit of the Stueckelberg theory\cite{Stuck} . The Stueckelberg theory is a theory of massive vector field, with an additional gauge-fixing term. A useful summary of this theory is provided by Ruegg and Ruiz-Altaba\cite{Stuckreview} .
The massless limit of the Stueckelberg theory, with finite Stueckelberg parameter \(\xi\), is equivalent to the massless vector theory, with covariant gauge-fixing term.

It is well known that the large-distance behaviour of the massless vector propagator in de Sitter spacetime is problematic\cite{MassLandauZeroNfour} \cite{MassLandauZero} , by which we mean that, for points with large (spacelike or timelike) separations, correlation functions do not decrease. It has been shown previously that the vector propagator tends to a non-zero constant in 4 dimensions, for \(\xi \neq 0\) \cite{MassLandauZeroNfour}. We show that the propagator behaves in the same way for general \(n\) dimensions, for \(n \geq 4\), in order to verify that this behaviour is not unique to 4 dimensions. Knowledge of the long-distance behaviour of the vector propagator in \(n\) dimensions will be useful in calculations involving dimensional regularisation. 

This same infrared pathology exists for the massless minimally coupled scalar field, as well as the graviton field. A method similar to the one 
used to calculate the long-distance behaviour of the massless vector propagator could be used to find the long-distance behaviour of the graviton two-point function, although it is expected that there will be additional difficulties associated with this calculation. For example, we will have to work to higher orders of the de Sitter-invariant distance \(Z\), 
as the graviton two-point function exhibits a linear divergence in \(Z\). Additionally, the graviton two-point function, found by Fr\"ob, Higuchi, and Lima\cite{masslessgravprop} , for example, has a larger number of terms than the vector two-point function. Both of these facts mean that the calculation of the large-distance behaviour of the graviton two-point function becomes far more lengthy than the calculation carried out in this paper.

The paper is structured as follows. In section 2 we give the massless vector propagator, and see that it is formed of three terms, which depend on the de Sitter-invariant distance \(Z\). 
As all terms are proportional to spacetime derivatives of \(Z\), in section 3 we present these derivatives. In section 4 we find the limit of the first two terms of the propagator, as the method involved in their calculation is very similar. In section 5 we calculate the final term. Finally, in section 6, we combine the results of the earlier sections to find an expression for the large-distance behaviour of the propagator. In this section we also show that our result agrees with the result found by Youssef\cite{MassLandauZeroNfour} for \(n=4\). We use the mostly plus convention for the metric.

\section{Preliminaries} \label{Prelim}
We consider the Stueckelberg theory\cite{Stuck} , which is an extension of the Proca theory, with an additional gauge-fixing term. We give a brief description of the Stueckelberg theory in this section, and further details necessary for the calculation will be provided throughout the paper. The Stueckelberg theory for massive vector fields would usually include a scalar of the same mass, which transforms under a gauge transformation such that we have a gauge symmetry, even in the massive case \cite{Stuckback1} 
\cite{Stuckback2} . In the massless case, the vector field alone possesses a gauge symmetry, so we ignore the scalar field in what follows. 

The Stueckelberg Lagrangian for a massive vector field is given by
\begin{equation}
\mathcal{L} = - \frac{1}{4} \sqrt{-g} \left[ F^{ab} F_{ab} + 2m^2 A_b A^b + \frac{2}{\xi} \left( \nabla_a A^a \right)^2 \right] \eqend{,}
\end{equation}
where the field strength tensor \(F_{ab} = \nabla_a A_b - \nabla_a A_b \), and \(\xi \) is the Stueckelberg parameter. For example, \(\xi=0\) corresponds to the Landau gauge, and \(\xi \to \infty \) gives the unitary gauge. We take the massless limit of this theory, which, as discussed in the introduction, corresponds to the standard massless vector theory, with a covariant gauge-fixing term. We work in de Sitter spacetime, described by the metric
\begin{equation}
\mathrm{d} s^2 = \frac{1}{(-H \eta)^2} \left( - \mathrm{d} \eta^2 + \mathrm{d} \boldsymbol{x} ^2 \right) \eqend{,}
\end{equation}
where \(H\) is a positive constant, and the conformal time \( \eta \in (-\infty , 0) \).  

In the following we use the notation used by Allen and Jacobson\cite{vectortwopoint} . By definition, for de Sitter,
\begin{equation}
Z (x,x') = \cos \left [ H \sigma (x,x') \right] \eqend{,} \label{Zdef}
\end{equation} 
where \(\sigma (x,x') \) is the spacelike geodesic distance between two-points \(x\) and \(x'\). If there is no geodesic connecting the points \(x\) and \(x'\), \(Z\) can be defined by analytic continuation. 
By definition, the parallel propagator is the unique solution to\cite{parallelprop}  
\begin{align}
\nabla^\rho \sigma (x,x') \nabla_\rho g_{\alpha \beta '} (x,x') = 0 \eqend{,}
\end{align}
where 
\begin{equation}
\lim_{x' \to x} g_{\alpha \beta '} (x,x') = g_{\alpha \beta} (x,x) \eqend{.}
\end{equation}
We also have
\begin{align}
g_{ab} (x) =& g_a^{c'} (x,x') g_{c'b} (x',x) \eqend{,} \\
g_{a'b'} (x') =& g_{a'}^c (x',x) g_{cb'} (x,x') \eqend{.}
\end{align}
Indices corresponding to \(x\) are unprimed, and indices corresponding to \(x'\) are primed. In most cases arguments are therefore omitted, as indices make it clear whether we refer to \(x\) or \(x'\). We raise / lower indices using the metrics \(g_{ab}\) and \(g_{a'b'}\) for unprimed and primed indices, respectively.  Additionally, we have
\begin{align}
n_a =& \nabla_a \sigma (x,x') \eqend{,} \\ 
n_{b'} =& \nabla_{b'} \sigma (x,x') \eqend{,}
\end{align}
where \(n_a\) and \(n_{b'}\) are unit vectors tangent to the geodesic at the points \(x\) and \(x'\) respectively. These vectors point away from each other, so 
\begin{equation}
g^{ab'} n_a = - n^{b'} \eqend{.}
\end{equation} 

We will find an expression for the vector two-point function in the limit where the points \(x\) and \(x'\) have a large separation in space or time coordinates. This is equivalent to the limit \(Z\to\pm\infty\). As is known, \(n\)-dimensional de Sitter spacetime can be considered to be the set of points \(X\) in \((n+1)\)-dimensional Minkowski spacetime such that \(X \cdot X =1\). We therefore work in conformal coordinates, and, for two points \(X =(\eta, \bx) \) and \(X'=(\eta' ,\bx ')\), define \(Z\) as 
\begin{equation}
Z = X \cdot X' = 1 - \frac{ || \boldsymbol{x} - \boldsymbol{x}'||^2  - (\eta - \eta ')^2}{2\eta \eta '} \eqend{.}
\end{equation}
Indeed, \(Z \to \pm \infty \) as the time- or space-coordinate separation between \(x\) and \(x'\) tends to infinity. 

The massless limit of the vector two-point function, as found by Fr\"ob and Higuchi\cite{MassLandauZero} , is 
\begin{align} \label{propagator}
\lim_{m^2 \to 0} \langle 0| A_a (x) A_{b'} (x') |0 \rangle =& \frac{H^{n-4}}{(4 \pi)^{\frac{n}{2}}} \bigg[ \frac{n-2}{n-3} I^{(0)} (Z) \partial_a \partial_{b'} Z + \frac{1}{n-3} I^{(0) \prime } (Z) (\partial_a Z) (\partial_{b'} Z) \bigg] \nonumber  \\
&+ \left( \xi - \frac{n-1}{n-3} \right) \partial_a \partial_{b'} \tilde{\laplace} (Z)  \eqend{,}
\end{align}
where
\begin{equation} \label{Izero}
I^{(0)} (Z) = \frac{\Gamma (n-2)}{\Gamma \left( \frac{n}{2} \right) } \text{ } _2 F_1 \left( n-2,1; \frac{n}{2} ; \frac{1+Z}{2} \right) \eqend{,}
\end{equation}
and \(_2 F_1 (a,b;c;z) \) is the hypergeometric function.
Additionally, we have defined
\begin{equation} \label{tildetriangle}
\tilde{\laplace} (Z) = - \lim_{m^2 \to 0} \frac{\partial }{\partial m^2} \left( \laplace_{m^2} (Z) -  \laplace_{m^2} (-1) \right) \eqend{,}
\end{equation}
where
\begin{equation}
\laplace_{m^2} (Z) = \frac{H^{n-2}}{(4 \pi)^{\frac{n}{2}}} I_\mu (Z) \eqend{,}
\end{equation}
and
\begin{equation} \label{Imu}
I_\mu (Z) = \frac{\Gamma \left( \frac{n-1}{2} + \mu \right) \Gamma \left( \frac{n-1}{2} - \mu \right)}{\Gamma \left( \frac{n}{2} \right)} \text{ }_2 F_1 \left( \frac{n-1}{2} + \mu , \frac{n-1}{2} - \mu ; \frac{n}{2} ; \frac{1+Z}{2} \right) \eqend{,}
\end{equation}
for
\begin{equation} \label{mu}
\mu = \sqrt{ \frac{(n-1)^2}{4} - \frac{m^2}{H^2} } \eqend{.}
\end{equation}
We here note that \(\xi=0\) corresponds to the massless vector two-point function found by Tsamis and Woodard\cite{xi0massvecprop} , and the massless vector propagator for \(\xi=1\), when \(n=4\), 
was given by Allen and Jacobson\cite{vectortwopoint} . 

In order to find the \(|Z| \to \infty \) limit of the propagator, defined by equation \eqref{propagator}, we find the large \(Z\) behaviour of each of the terms in the propagator separately, before combining the results at the end. The derivatives of \(Z\) appear in all terms, so we start by finding an explicit expression for these.

\section{Derivatives of \(Z\)} \label{ZDeriv}
The derivatives are calculated by Allen and Jacobson\cite{vectortwopoint} , and we present them here for later use. No approximations are made in this calculation, so the results of this section are exact. 

From the definition of \(Z\), given by equation \eqref{Zdef}, one obtains
\begin{equation}
\partial_{b'} Z = - H \sqrt{1-Z^2} n_{b'} \eqend{,}  \label{Zprime}
\end{equation}
so
\begin{equation}
(\partial_a Z) (\partial_{b'} Z) = H^2 (1-Z^2) n_a n_{b'} \eqend{,} \label{ZprimeZprime}
\end{equation}
and
\begin{equation}
\partial_a \partial_{b'} Z= H^2 \left[ g_{ab'} + (1-Z) n_a n_{b'} \right] \eqend{,} \label{Zprimeprime}
\end{equation}
where the results of Allen and Jacboson\cite{vectortwopoint} have been used to evaluate \( \partial_{b'} n_a\). The results \eqref{ZprimeZprime} and \eqref{Zprimeprime} are frequently referred to in the rest of this paper. 

The final term of the propagator, given by equation \eqref{propagator}, is \(\partial_a \partial_{b'} \tilde{\laplace} (Z)\). We now use the chain rule, along with the results of this section, to express this as
\begin{align}
\partial_a \partial_{b'} \tilde{\laplace} (Z) &= (\partial_a \partial_{b'} Z )  \tilde{\laplace} ' (Z) + ( \partial_{b'} Z ) ( \partial_a Z ) \tilde{\laplace} ^{\prime \prime} (Z) \eqend{,} \nonumber \\
&=  H^2 \left[ g_{ab'} + (1-Z) n_a n_{b'} \right] \tilde{\laplace} ' (Z) + H^2 (1-Z^2) n_a n_{b'} \tilde{\laplace} ^{\prime \prime} (Z) \eqend{.} \label{chainrule}
\end{align}

The tensor structure of the propagator ensures that each term is proportional to either \(g_{ab'}\) or \(n_a n_{b'}\). We therefore write the propagator as
\begin{equation} \label{propagatorAB}
\lim_{m^2 \to 0} \langle 0| A_a (x) A_{b'} (x') |0 \rangle  = A (Z) g_{ab'} + B (Z) n_a n_{b'} \eqend{,}
\end{equation}
where 
\begin{equation} \label{A}
A(Z) = \frac{H^{n-2}}{(4 \pi)^{\frac{n}{2}}} \frac{n-2}{n-3} I^{(0)} (Z) + H^2 \left( \xi - \frac{n-1}{n-3} \right) \tilde{\laplace} ' (Z) \eqend{,}
\end{equation}
and 
\begin{align} \label{B}
B(Z) =& \frac{H^{n-2}}{(4 \pi)^{\frac{n}{2}}} \bigg[ \frac{n-2}{n-3} (1-Z) I^{(0)} (Z) + \frac{1}{n-3} (1-Z^2) I^{(0) \prime } (Z) \bigg] \nonumber \\
&+ H^2 \left( \xi - \frac{n-1}{n-3} \right) \left[  (1-Z)  \tilde{\laplace} ' (Z) + (1-Z^2) \tilde{\laplace} ^{\prime \prime} (Z) \right] \eqend{.}
\end{align}
The leading order behaviour of \(A(Z)\) and \(B(Z)\) is different; we therefore find \(A (Z)\) and \(B (Z)\) separately.

\section{Large-distance limit of the first two terms of the propagator} \label{term12}
The \(|Z| \to \infty \) limit of the first two terms in the propagator, equation \eqref{propagator}, are found using a very similar method; we therefore consider these two terms together. As we have already found \( \partial_a \partial_{b'} Z \) and \( (\partial_a Z) (\partial_{b'} Z) \), it only remains to find the large \(Z\) behaviour of \(I^{(0)} (Z)\) and \(I^{(0)\prime} (Z)\), which is the focus of this section.

\subsection{Calculation of the large-distance limit of \(I^{(0)} (Z)\)}
\(I^{(0)} (Z)\), as defined in equation \eqref{Izero}, is proportional to the hypergeometric function. In order to study the behaviour of \(I^{(0)} (Z)\) in the limit \(|Z|\to \infty\), we use the well known transformation property\cite{HypeGeom} , 
\begin{align} \label{hypegeomtrans}
_2 F_1 (a,b;c;z) =& \frac{\Gamma(c) \Gamma(b-a)}{\Gamma(b)\Gamma(c-a)} (1-z)^{-a} \text{ } _2 F_1 \left( a,c-b;a-b+1;\frac{1}{1-z} \right) \nonumber \\
&+ \frac{\Gamma(c) \Gamma(a-b)}{\Gamma(a)\Gamma(c-b)} (1-z)^{-b} \text{ } _2 F_1 \left( b,c-a;b-a+1;\frac{1}{1-z} \right) \eqend{.}
\end{align}
Using this transformation on equation \eqref{Izero}, and expanding for large \(Z\), we see that  
\begin{align}
I^{(0)} (Z) \to  \frac{ \Gamma (n-3) }{\Gamma \left( \frac{n}{2} - 1\right)} \left( \frac{1-Z}{2} \right)^{-1} + \mathcal{O} \left( Z^{-2} \right) \eqend{.}  \label{Izerolimit}
\end{align}
The transformation given by equation \eqref{hypegeomtrans} gives us a sum of two hypergeometric functions. Our leading order behaviour comes from the leading order term of the second hypergeometric function. The leading order contribution from the first hypergeomtric function is
\begin{equation} \label{singular}
\frac{\Gamma (n-2) \Gamma (3-n) }{\Gamma \left( 2- \frac{n}{2} \right)} \left( \frac{1-Z}{2} \right)^{-(n-2)}  \eqend{,}
\end{equation}
which, for \(n>2\), is of lower order than \((1-Z)^{-2}\). As we consider \(n \geqslant 4\), we are free to ignore the first hypergeometric function from the transformation given by equation \eqref{hypegeomtrans}\footnote[1]{Details of \(n=2,3,4\) cases are described by Fr\"ob and Higuchi\cite{MassLandauZero}.}. 
We note that, despite the apparent singularity due to the integer \(n\) case of the \(\Gamma(3-n)\) factor in equation \eqref{singular}, the result of this section is well-defined\footnote[2]{The singular terms combine with singular terms which occur at higher orders in the second line of equation \eqref{Izerolimit}, to give a well-defined contribution in the \(n \to\) integer limit. This term behaves like \([(1-Z)/2]^{-(n-2)}\ln Z\), which is still suppressed in comparison with \([(1-Z)/2]^{-1}\). }.

It can be seen that neglecting all other terms other than the leading order behaviour of equation \eqref{Izerolimit} is valid, as the neglected terms have no effect on the large-distance behaviour of the propagator. 
This can be seen most clearly from the explicit expressions for \(A(Z)\) and \(B(Z)\), given by equations \eqref{A} and \eqref{B}, respectively. In \(A(Z)\), \(I^{(0)}(Z)\) appears without being multiplied by any function of \(Z\), so the contribution to the propagator from all terms in equation \eqref{Izerolimit} vanish in the large \(Z\) limit. In \(B(Z)\), \(I^{(0)}(Z)\) appears when multiplied by a factor of \((1-Z)\), so only the contribution from the first term in equation \eqref{Izerolimit} is non-vanishing in the limit \(|Z| \to \infty\). 



Therefore, in the \(|Z| \to \infty\) limit, we find, 
\begin{equation} \label{Inought}
I^{(0)} (Z) \partial_a \partial_{b'} Z \to 2H^2 \frac{\Gamma (n-3)}{\Gamma \left( \frac{n}{2} -1 \right)} n_a n_{b'} + \mathcal{O} \left( Z^{-1} \right) \eqend{,}
\end{equation}
which, up to a constant, gives the \(|Z| \to \infty \) limit of the first term of the propagator. 

\subsection{Calculation of the large-distance limit of \(I^{(0) \prime}(Z)\)}
In order to calculate \(I^{(0) \prime}(Z)\), first note that the derivative of the hypergeometric function is
\begin{equation}
\frac{\mathrm{d}}{\mathrm{d} z} \text{ } _2F_1(a,b;c;z) = \frac{ab}{c} \text{ } _2F_1 (a+1,b+1;c+1;z) \eqend{.}
\end{equation}
The derivative of \(I^{(0)}(Z)\), defined by \eqref{Izero}, is therefore
\begin{align}
I^{(0) \prime } (Z) &= \frac{\mathrm{d}}{\mathrm{d} Z} I^{(0)} (Z) \eqend{,} \nonumber \\
&= \frac{\Gamma (n-1)}{n\Gamma \left( \frac{n}{2} \right) } \text{ } _2F_1 \left( n-1,2;\frac{n}{2}+1;\frac{1+Z}{2} \right) \eqend{.}
\end{align}
Using the same transformation of the hypergeometric function as in the previous subsection, given by equation \eqref{hypegeomtrans}, we see that 
\begin{equation}
I^{(0) \prime } (Z) \to  \frac{\Gamma(n-3)}{2 \Gamma \left( \frac{n}{2} -1 \right)} \left( \frac{1-Z}{2} \right)^{-2} + \mathcal{O} \left( Z^{-3} \right) \eqend{.} \label{IprimeZerolimit}
\end{equation}
As in the previous section, the contribution to equation \eqref{IprimeZerolimit} from the first hypergeometric function in the transformation, given by equation \eqref{hypegeomtrans}, can be neglected. To see this, we note that the leading order term in the expansion of the first hypergeometric function in equation \eqref{hypegeomtrans} is\footnote[3]{As in the previous section, the integer \(n\) singularity from the \(\Gamma (3-n)\) factor in the following equation is cancelled by higher order contributions from the expansion of the second hypergeometric function in equation \eqref{hypegeomtrans}} 
\begin{equation}
\frac{\Gamma (n-1) \Gamma(3-n)}{2\Gamma \left( 2- \frac{n}{2}\right)}  \left( \frac{1-Z}{2} \right)^{-(n-1)} \eqend{,}
\end{equation}
which, for \(n>3\), is of higher order in \(\frac{1}{1-Z}\) than the leading order behaviour of \eqref{IprimeZerolimit}. As quick check we note that the same result can be attained by differentiating the large \(Z\) limit of \(I^{(0)} (Z)\), found in equation \eqref{Izerolimit} in the previous subsection. 

As in the calculation of \(I^{(0)} (Z)\), we verify that neglecting all but the leading order term in equation \eqref{IprimeZerolimit} has no effect on the large-distance behaviour of the propagator. To this end, we note that \(I^{(0) \prime } (Z)\) only appears in the propagator in the function \(B(Z)\), given by equation \eqref{B}. It appears here multiplied by a factor of \((1-Z^2)\), the contribution of \(I^{(0) \prime } (Z)\) to \(B(Z)\) comes only from the leading order term in equation \eqref{IprimeZerolimit}, so we are free to neglect higher orders of \( \frac{1}{1-Z} \). 

For large \(Z\), we therefore see that
\begin{equation} \label{Inoughtprime}
I^{(0) \prime } (Z) (\partial_a Z)(\partial_{b'}Z) \to -H^2 \frac{\Gamma(n-3)}{\Gamma \left( \frac{n}{2} -1 \right)} n_a n_{b'} + \mathcal{O} \left( Z^{-1} \right)  \eqend{,}
\end{equation}
as, in the limit \(|Z| \to \infty\),
\begin{equation}
\frac{1-Z^2}{(1-Z)^2} = \frac{1+Z}{1-Z} \to -1 + \mathcal{O} \left( Z^{-1} \right) \eqend{.}
\end{equation}
Equation \eqref{Inoughtprime}, gives, up to an \(n\) dependent constant factor, the large-distance behaviour of the second term in the propagator. 

\subsection{Summary}
We combine the results of the last two subsections to find the limit as \(|Z| \to \infty \) of the first line of the propagator. Substituting equations \eqref{Inought} and \eqref{Inoughtprime} into the first line of equation \eqref{propagator} gives 
\begin{align}
\frac{H^{n-2}}{(4\pi)^{\frac{n}{2}}} \frac{\Gamma (n-1)}{\Gamma \left( \frac{n}{2} \right) } \frac{1}{n-3} n_a n_{b'} + \mathcal{O} \left( Z^{-1} \right)  \eqend{,} \label{firsttwoterms}
\end{align}
where the prefactor involving gamma functions has been written in a slightly different form, for later convenience.

\section{Large-distance limit of the final term of the propagator}
A different approach must be used to find the large \(Z\) behaviour of the final term in the propagator. Specifically, we find the limit as \(|Z| \to \infty\) of 
\begin{equation}
\tilde{\laplace} (Z) = - \frac{H^{n-2}}{(4 \pi)^{\frac{n}{2}}} \lim_{m^2 \to 0} \frac{\partial }{\partial m^2} \left( I_\mu (Z) -  I_\mu (-1) \right) \eqend{,} \label{laplacetilde}
\end{equation}
where
\begin{align}
I_\mu (Z) =& \frac{\Gamma \left( \frac{n-1}{2} + \mu \right) \Gamma \left( \frac{n-1}{2} - \mu \right)}{\Gamma \left( \frac{n}{2} \right)} \text{ } _2F_1 \left( \frac{n-1}{2} + \mu , \frac{n-1}{2} - \mu ; \frac{n}{2} ; \frac{1+Z}{2} \right) \eqend{,} \\
\mu =& \sqrt{ \frac{(n-1)^2}{4} - \frac{m^2}{H^2} } \eqend{,}
\end{align}
as defined in section \ref{Prelim}, repeated here for convenience. In section 3, we wrote \(\tilde{\laplace} (Z)\) in terms of the functions \(\tilde{\laplace} '(Z) \) and \(\tilde{\laplace}''(Z)\), and derivatives of \(Z\). For convenience, we repeat this expression here: 
\begin{align}
\partial_a \partial_{b'} \tilde{\laplace} (Z) &= (\partial_a \partial_{b'} Z )  \tilde{\laplace} ' (Z) + ( \partial_{b'} Z ) ( \partial_a Z ) \tilde{\laplace} ^{\prime \prime} (Z) \eqend{,} \nonumber \\
&=  H^2 \left[ g_{ab'} + (1-Z) n_a n_{b'} \right] \tilde{\laplace} ' (Z) + H^2 (1-Z^2) n_a n_{b'} \tilde{\laplace} ^{\prime \prime} (Z) \eqend{.} 
\end{align}
It therefore only remains to find the large-distance behaviour of the functions \(\tilde{\laplace} '(Z) \) and \(\tilde{\laplace}''(Z)\), which is the focus of this section.

\subsection{Calculation of the large-distance limit of the first and second derivatives of \(\tilde{\laplace}(Z)\)}
The functions \(\tilde{\laplace} '(Z) \) and \(\tilde{\laplace}''(Z)\) are calculated using the same method, so we consider both functions at once. To find \( \tilde{\laplace} '(Z) \) and \(\tilde{\laplace}''(Z)\), we first differentiate \(\tilde{\laplace} (Z)\), defined in equation \eqref{laplacetilde}, once with respect to \(Z\), to find that
\begin{eqnarray}
\tilde{\laplace} ' (Z) &=& - \frac{H^{n-2}}{(4 \pi)^{\frac{n}{2}}} \lim_{m^2 \to 0} \frac{\partial }{\partial m^2} \left( \frac{\mathrm{d}}{\mathrm{d} Z} I_\mu (Z) -  \frac{\mathrm{d}}{\mathrm{d} Z} I_\mu (-1) \right) \eqend{,} \nonumber \\
&=& - \frac{H^{n-2}}{2(4 \pi)^{\frac{n}{2}}} \lim_{m^2 \to 0} \frac{\partial }{\partial m^2} \Bigg[ \frac{\Gamma \left( \frac{n+1}{2} + \mu \right) \Gamma \left( \frac{n+1}{2} - \mu \right)}{\Gamma \left( \frac{n}{2} +1 \right)} \nonumber \\
&& \ \ \ \ \ \ \ \ \ \ \ \ \ \ \ \ \ \ \ \ \ \ \ \ \ \ \ \ \ \ \times \text{ } _2F_1 \left( \frac{n+1}{2} + \mu , \frac{n+1}{2} - \mu ; \frac{n}{2} +1 ; \frac{1+Z}{2} \right) \Bigg]\eqend{,} \nonumber \\
\label{onederiv}
\end{eqnarray}
where elementary properties of the gamma function are used to simplify the prefactor. The second term in the first line of the latter equation, proportional to \(I_\mu (-1)\), vanishes, as this has no dependence on \(Z\). Similarly, we see that
\begin{eqnarray}
\tilde{\laplace} '' (Z) &=& - \frac{H^{n-2}}{(4 \pi)^{\frac{n}{2}}} \lim_{m^2 \to 0} \frac{\partial }{\partial m^2}\left( \frac{\mathrm{d}^2}{\mathrm{d} Z^2} I_\mu (Z) -  \frac{\mathrm{d}^2}{\mathrm{d} Z^2} I_\mu (-1) \right) \eqend{,} \nonumber \\
&=& - \frac{H^{n-2}}{4(4 \pi)^{\frac{n}{2}}} \lim_{m^2 \to 0} \frac{\partial }{\partial m^2} \Bigg[ \frac{\Gamma \left( \frac{n+3}{2} + \mu \right) \Gamma \left( \frac{n+3}{2} - \mu \right)}{\Gamma \left( \frac{n}{2} +2 \right)} \nonumber \\
&& \ \ \ \ \ \ \ \ \ \ \ \ \ \ \ \ \ \ \ \ \ \ \ \ \ \ \ \ \ \ \times \text{ } _2F_1 \left( \frac{n+3}{2} + \mu , \frac{n+3}{2} - \mu ; \frac{n}{2} +2 ; \frac{1+Z}{2} \right) \Bigg] \eqend{.} \nonumber \\
\label{twoderiv}
\end{eqnarray}
In order to find \(\tilde{\laplace} ' (Z)\) and \(\tilde{\laplace} '' (Z)\), given by equations \eqref{onederiv} and \eqref{twoderiv} respectively, we first find the large \(Z\) limit of the hypergeometric function, which appears as a factor in each case. We then find the derivative with respect to \(m^2\) of this, before taking the limit \(m^2 \to 0\). As the hypergeometric function is 
analytic in its parameters, this is a valid approach to take.

To find the large \(Z\) limit of \(\tilde{\laplace} '(Z)\) and \(\tilde{\laplace} ''(Z)\), we first apply the transformation of the hypergeometric function used in previous sections, given by equation \eqref{hypegeomtrans}. This gives 
\begin{align} \label{onederivtrans}
\tilde{\laplace} '(Z) \to & - \frac{H^{n-2}}{2(4 \pi)^{\frac{n}{2}}} \lim_{m^2 \to 0} \frac{\partial }{\partial m^2} \frac{\Gamma \left( \frac{n+1}{2} - \mu \right) \Gamma (2\mu)}{\Gamma \left( \frac{1}{2} +\mu \right)} \left( \frac{1-Z}{2} \right)^{-(\frac{n+1}{2} - \mu)} \nonumber \\
&+ \mathcal{O} \left( Z^{-(\frac{n+3}{2} - \mu)} \right) \eqend{.}
\end{align} 
This is the leading order term in the expansion of the second hypergeometric function in equation \eqref{hypegeomtrans}. In the limit \(m^2 \to 0\) \footnote[4]{When \(m^2 \to 0\), \(\mu \to \frac{n-1}{2}\).}, the large \(Z\) expansion of the first hypergeometric function in equation \eqref{hypegeomtrans} would give a leading term of order \( (1-Z)^{-n} \),  which, for \(n>2\) can be neglected. Similarly, 
\begin{align} \label{twoderivtrans}
\tilde{\laplace} ''(Z) \to & - \frac{H^{n-2}}{4(4 \pi)^{\frac{n}{2}}} \lim_{m^2 \to 0} \frac{\partial }{\partial m^2} \frac{\Gamma \left( \frac{n+3}{2} - \mu \right) \Gamma (2\mu)}{\Gamma \left( \frac{1}{2} +\mu \right)} \left( \frac{1-Z}{2} \right)^{-(\frac{n+3}{2} - \mu)} \nonumber \\
&+ \mathcal{O} \left( Z^{-(\frac{n+5}{2} - \mu)} \right)  \eqend{.}
\end{align}
Again, this term is the leading order term in the expansion of the second hypergeometric function in equation \eqref{hypegeomtrans}. In the large \(Z\), and \(m^2 \to 0 \) limits, the expansion of the first hypergeometric function in equation \eqref{hypegeomtrans} gives a leading term of order \( (1-Z)^{-(n+1)}  \), which can also be neglected when \(n>2\). As we look in the case when \(n>4\), we work to the order \(Z\) given in equations \eqref{onederivtrans} and \eqref{twoderivtrans}. 
Equations \eqref{onederivtrans} and \eqref{twoderivtrans} have the same basic structure. For ease of notation, we therefore define
\begin{equation}
\chi = \lim_{m^2 \to 0} \frac{\partial }{\partial \mu} \left[ f(\mu) \left( \frac{1-Z}{2} \right)^{-g(\mu)}  \right] \frac{\partial \mu}{\partial m^2} \eqend{,} 
\label{lfgequation}
\end{equation}
where
\begin{align}
f(\mu) &= \frac{\Gamma(l-\mu) \Gamma(2\mu)}{\Gamma \left( \frac{1}{2} + \mu \right) } \eqend{,} \label{fdef}  \\
g(\mu) &= l-\mu \eqend{.} \label{gdef}
\end{align}
Equation \eqref{lfgequation} gives the leading order behaviour of equations \eqref{onederivtrans} and \eqref{twoderivtrans}, for \(l= \frac{n+1}{2}\) and \(l= \frac{n+3}{2} \), respectively, up to an overall multiplicative constant. From the definition of \(\mu\), given by \eqref{mu},
\begin{equation}
\frac{\partial \mu}{\partial m^2} = -\frac{1}{2H^2 \mu} \eqend{,}
\end{equation}
and we see that, as \(m^2 \to 0\), \(\mu \to \frac{n-1}{2}\). The function \(\chi\), given by equation \eqref{lfgequation}, becomes
\begin{equation}
\chi = -\frac{1}{H^2 (n-1)} \lim_{\mu \to \frac{n-1}{2}} \left[ \Bigg( f'(\mu) -f(\mu) g'(\mu) \log \left( \frac{1-Z}{2} \right) \Bigg) \left( \frac{1-Z}{2} \right)^{-g(\mu)} \right] \eqend{.} \label{chimid}
\end{equation}
It now remains to find \(f'(\mu)\) and \(g'(\mu)\). As \(f(\mu)\) consists of gamma functions, we start by noting that the derivative of the gamma function is given by 
\begin{equation}
\Gamma '(x) = \Gamma (x) \psi_0 (x) \eqend{,}
\end{equation}
where the digamma function \(\psi_0(n)\) has the following value at \(x=n \in \mathbb{N}\),
\begin{equation} \label{digamma}
\psi_0 (n) = -\gamma + \sum_{k=1}^{n-1} \frac{1}{k} \eqend{,}
\end{equation}
and, for a half-integer argument, 
\begin{equation}
\psi_0 \left( n+ \frac{1}{2} \right) = -\gamma - 2 \ln 2 + \sum_{k=1}^n \frac{2}{2k-1} \eqend{,}
\end{equation}
where \(\gamma \) is the Euler-Mascheroni constant. We therefore find that
\begin{align}
f' (\mu) &= \left[ \frac{-\Gamma '(l-\mu) \Gamma(2\mu) + 2 \Gamma(l-\mu) \Gamma '(2\mu)}{\Gamma \left( \frac{1}{2} + \mu \right) } - \frac{\Gamma (l-\mu) \Gamma(2\mu) \Gamma' \left( \frac{1}{2} +\mu \right) }{ \left( \Gamma \left( \frac{1}{2} +\mu \right) \right)^2} \right] \eqend{,} \nonumber \\
&= - f(\mu) Q(l) \eqend{,} \label{fprimemu}
\end{align} 
where we define
\begin{equation} \label{Qdef}
Q (l) = \psi_0 \left( l - \frac{n-1}{2} \right) - 2 \psi_0 (n-1) + \psi_0 \left( \frac{n}{2} \right) \eqend{.}
\end{equation}
We also find, trivially, that
\begin{equation}
g'(\mu) = -1 \eqend{.} \label{gprimemu}
\end{equation}
By evaluating our functions \(f(\mu)\), \(g(\mu)\), \(f'(\mu)\), and \(g'(\mu)\), given by equations \eqref{fdef}, \eqref{gdef}, \eqref{fprimemu}, and \eqref{gprimemu}, respectively, at \(\mu = \frac{n-1}{2}\), and substituting into equation \eqref{chimid}, we see that
\begin{align}
\chi =\frac{1}{H^2 (n-1)} \frac{\Gamma(l-\frac{n-1}{2}) \Gamma(n-1)}{\Gamma \left( \frac{n}{2} \right) } \left[ Q(l) - \log \left( \frac{1-Z}{2} \right) \right] \left( \frac{1-Z}{2} \right)^{-( l- \frac{n-1}{2} )} \eqend{.} \label{chifinal}
\end{align}
We can now use this expression for \(\chi\), given by \eqref{chifinal}, to find \(\tilde{\laplace} '(Z)\) and \(\tilde{\laplace} ''(Z)\), as defined in \eqref{onederivtrans} and \eqref{twoderivtrans} respectively. Remembering to include the \(n\) dependent constant prefactor, and to put in the corresponding values of \(l\), we find
\begin{align}
\tilde{\laplace} '(Z) \to & - \frac{H^{n-4}}{(4\pi)^{\frac{n}{2}} (n-1)} \frac{\Gamma (n-1)}{\Gamma \left( \frac{n}{2} \right)} \Bigg[ Q \left( \frac{n+1}{2} \right) - \log \left( \frac{1-Z}{2} \right) \Bigg] \frac{1}{1-Z} \nonumber \\
&+ \mathcal{O} \left( Z^{-2} \right) \eqend{,} \label{onederivresult}
\end{align}
and
\begin{align}
\tilde{\laplace} ''(Z) \to & - \frac{H^{n-4}}{(4\pi)^{\frac{n}{2}} (n-1)} \frac{\Gamma (n-1)}{\Gamma \left( \frac{n}{2} \right)} \Bigg[ Q \left( \frac{n+3}{2} \right) - \log \left( \frac{1-Z}{2} \right) \Bigg] \frac{1}{(1-Z)^2} \nonumber \\
&+ \mathcal{O} \left( Z^{-3} \right) \eqend{.} \label{twoderivresult} 
\end{align}
To see why we work to these orders in \(\tilde{\laplace} '(Z)\) and \(\tilde{\laplace} ''(Z)\), we refer back to the expression of the propagator, given by equation \eqref{propagatorAB}. As the function \(\tilde{\laplace} '(Z)\) appears in the propagator through \(A(Z)\), and through \(B(Z)\) when multiplied by \((1-Z)\), all non-leading order contributions to the propagator from equation \eqref{onederivresult} vanish, so we are free to neglect these terms. The function \(\tilde{\laplace} ''(Z)\) only appears in the propagator through \(B(Z)\), where it is multiplied by \((1-Z^2)\). The only contribution to the propagator from \(\tilde{\laplace} ''(Z) \) therefore comes from this leading order term, so we are free to neglect terms of order \((1-Z)^{-3} \) and higher.  

These expressions for the large \(Z\) behaviour of \(\tilde{\laplace} '(Z)\) and \(\tilde{\laplace} ''(Z)\) can now be combined with the results of section \ref{ZDeriv} to find the large \(Z\) behaviour of \(\partial_a \partial_{b'} \laplace Z\), given by equation \eqref{chainrule}, which is, up to a gauge dependent constant, the final term of the propagator.

\subsection{Summary}
We now use equations \eqref{onederivresult} and \eqref{twoderivresult}, along with the results of section \ref{ZDeriv}, to see that, in the limit as \(|Z| \to \infty\), equation \eqref{chainrule} becomes
\begin{equation}
\partial_a \partial_{b'} \tilde{\laplace} (Z) \to \frac{H^{n-2} \Gamma(n-1) }{(4\pi)^{\frac{n}{2}} (n-1) \Gamma \left( \frac{n}{2} \right)} n_a n_{b'} + \mathcal{O} \left( Z^{-1} \right) \eqend{,} \label{lastterm}
\end{equation}
as
\begin{equation}
Q \left( \frac{n+1}{2} \right) - Q \left( \frac{n+3}{2} \right) = \psi_0 (1) - \psi_0 (2) = -1 \eqend{.}
\end{equation}
The latter follows from the definition of \(Q(l)\), given by equation \eqref{Qdef}, and the representation of the digamma function given by equation \eqref{digamma}.

We will now combine this with the result of section \ref{term12} to find the propagator in the \(|Z| \to \infty\) limit.

\section{Large-distance behaviour of the propagator}
We now combine the results of the previous sections to find an expression for the \(|Z| \to \infty \) limit of the propagator. As stated before, in section 3, the large-distance behaviour of the propagator, is 
\begin{equation}
\lim_{m^2 \to 0} \langle 0| A_a (x) A_{b'} (x') |0 \rangle  =  A(Z) g_{ab'} + B(Z) n_a n_{b'} \eqend{,}
\end{equation}
for \(A(Z)\) and \(B(Z)\) defined by equations \eqref{A} and \eqref{B}, respectively. 

As equations \eqref{firsttwoterms} and \eqref{lastterm} both have no terms proportional to \(g_{ab'}\), we find that
\begin{equation}
A(Z) \to \mathcal{O} \left( Z^{-1} \right) \eqend{,}
\end{equation}
as \(|Z| \to \infty\). Additionally, combining terms proportional to \(n_a n_{b'}\) in equations \eqref{firsttwoterms} and \eqref{lastterm}, we see that
\begin{equation}
B(Z) \to  \xi \frac{H^{n-2} \Gamma (n-1)}{(4\pi)^{\frac{n}{2}} \Gamma \left( \frac{n}{2} \right) (n-1)} +  \mathcal{O} \left( Z^{-1} \right) \eqend{.}
\end{equation}
as \(|Z| \to \infty\).

We therefore see that, in \(|Z| \to \infty \) limit, the propagator is proportional to \(n_a n_{b'}\), and tends to 
\begin{equation} \label{proplim}
\lim_{m^2 \to 0} \langle 0| A_a (x) A_{b'} (x') |0 \rangle  \to \xi \frac{H^{n-2} \Gamma (n-1)}{(4\pi)^{\frac{n}{2}} \Gamma \left( \frac{n}{2} \right) (n-1)} n_a n_{b'} + \mathcal{O} \left( Z^{-1} \right) \eqend{,}
\end{equation}
which is a non-zero gauge dependent constant. 
Hence, in the limit \(|Z| \to \infty \), the two-point function vanishes in the Landau gauge, but tends to a non-zero constant in a general gauge where \(\xi \neq 0\). Our result, given by equation \eqref{proplim}, 
reduces, for \(n=4\), to the result found by Youssef\cite{MassLandauZeroNfour} , which is that
\begin{equation}
\beta (Z) \to \frac{\xi H^2}{24 \pi^2} \eqend{.}
\end{equation}

\section{Discussion}
In this article we have found the large \(Z\) behaviour of the massless vector propagator in \(n\)-dimensional de Sitter space. 
As in the \(n=4\), case discussed by Youssef\cite{MassLandauZeroNfour} , it was found, in the large \(Z\) limit, that the propagator tends towards a gauge dependent constant. It was found that this constant is equal to zero in the Landau gauge, where \(\xi=0\). 

It is expected that a method similar to the one used in this paper can be used to find the large \(Z\) behaviour of the graviton two-point function, which is also known to experience problems in the infrared.

\section*{Acknowledgements}
I would like to thank Atsushi Higuchi for helpful discussion and a critical reading of the manuscript. This work was supported by the Engineering and Physical Sciences Research Council (EPSRC). 

\bibliographystyle{unsrt}
\bibliography{year1}

\end{document}